\documentclass[showpacs,preprintnumbers,superscriptaddress,amsmath,amssymb, pre]{revtex4-1}

\usepackage{graphicx,easymat,chemarrow}
\usepackage{dcolumn}
\usepackage{bm}
\usepackage{color}
\usepackage{multirow}
\usepackage{listings}
\usepackage{CJK}

\begin{document}
\begin{CJK*}{GBK}{Song} 

\preprint{FoP/MF-X-WL}

\title{Joint multifractal analysis based on wavelet leaders}

\author{Zhi-Qiang Jiang}
\email{zqjiang@ecust.edu.cn}
 \affiliation{School of Business, East China University of Science and Technology, Shanghai 200237, China}
 \affiliation{Research Center for Econophysics, East China University of Science and Technology, Shanghai 200237, China}
 \affiliation{Department of Physics and Center for Polymer Studies,
   Boston  University, Boston, MA 02215, USA}

 \author{Yan-Hong Yang}
 \affiliation{School of Business, East China University of Science and Technology, Shanghai 200237, China}
 \affiliation{Research Center for Econophysics, East China University of Science and Technology, Shanghai 200237, China}
 \affiliation{Department of Physics and Center for Polymer Studies,
   Boston  University, Boston, MA 02215, USA}

 \author{Gang-Jin Wang}
  \email{wanggangjin@hnu.edu.cn}
\affiliation{Department of Physics and Center for Polymer Studies, Boston  University, Boston, MA 02215, USA}
 \affiliation{Business School and Center of Finance and Investment Management, Hunan University, Changsha 410082, China} %

\author{Wei-Xing Zhou}
 \email{wxzhou@ecust.edu.cn}
 \affiliation{School of Business, East China University of Science and Technology, Shanghai 200237, China}
 \affiliation{Research Center for Econophysics, East China University of Science and Technology, Shanghai 200237, China}
 \affiliation{Department of Mathematics, East China University of Science and Technology, Shanghai 200237, China}


\date{\today}

\begin{abstract}
  Mutually interacting components form complex systems and the outputs of these components are usually long-range cross-correlated. Using wavelet leaders, we propose a method of characterizing the joint multifractal nature of these long-range cross correlations, a method we call joint multifractal analysis based on wavelet leaders (MF-X-WL). We test the validity of the MF-X-WL method by performing extensive numerical experiments on the dual binomial measures with multifractal cross correlations and the bivariate fractional Brownian motions (bFBMs) with monofractal cross correlations. Both experiments indicate that MF-X-WL is capable to detect the cross correlations in synthetic data with acceptable estimating errors. We also apply the MF-X-WL method to the pairs of series from financial markets (returns and volatilities) and online worlds (online numbers of different genders and different societies) and find an intriguing joint multifractal behavior.
\end{abstract}


\maketitle

\end{CJK*}

\section{Introduction}
\label{sec:Introduction}

Since the seminal paper on long-range cross-correlation analysis \cite{Podobnik-Stanley-2008-PRL}, the cross correlation and joint multifractality have received considerable research interests. Great concentrations are focused on extending the traditional multifractal detecting approaches into cross or joint multifractal formulism and applying such methods to diagnose the cross or joint multifractality in many real systems.

The invented cross or joint multifractal analysis methods are all rooted from the traditional approaches on multifractal analysis, such as partition function methods \cite{Grassberger-1983-PLA,Grassberger-Procaccia-1983-PD,Halsey-Jensen-Kadanoff-Procaccia-Shraiman-1986-PRA}, structure function methods\cite{Kolmogorov-1962-JFM,VanAtta-Chen-1970-JFM,Anselmet-Gagne-Hopfinger-Antonia-1984-JFM}, detrended fluctuation analysis \cite{CastroESilva-Moreira-1997-PA,Weber-Talkner-2001-JGR,Kantelhardt-Zschiegner-KoscielnyBunde-Havlin-Bunde-Stanley-2002-PA}, and detrending moving-average analysis \cite{Alessio-Carbone-Castelli-Frappietro-2002-EPJB,Carbone-Castelli-Stanley-2004-PA,Carbone-Castelli-Stanley-2004-PRE,Gu-Zhou-2010-PRE}, and so on. In 1990, Meneveau {\emph{et al.}} proposed a joint multifractal analysis to handle the joint partition function of two multifractal measures and to study the relationship between the dissipation rates of kinetic energy and passive scalar fluctuations in fully developed turbulence \cite{Meneveau-Sreenivasan-Kailasnath-Fan-1990-PRA}, which is also termed as the multifractal cross-correlation analysis based on the partition function approach (MF-X-PF) \cite{Xie-Jiang-Gu-Xiong-Zhou-2015-NJP}. Wang {\emph{et al.}} studied independently the multifractal statistical moment cross-correlation analysis (MFSMXA) \cite{Wang-Shang-Ge-2012-Fractals}, which is a special case of the MF-X-PF. Xie {\emph{et al.}} theoretically derived and numerically validated the expression of multifractal formula for binomial measures \cite{Xie-Jiang-Gu-Xiong-Zhou-2015-NJP}. Kristoufek proposed the multifractal height cross-correlation analysis (MF-HXA) based on structure function \cite{Kristoufek-2011-EPL}. Zhou generalized the detrended fluctuation analysis into multifractal detrended cross-correlation analysis (MF-X-DFA) \cite{Zhou-2008-PRE},  which is a multifractal version of detrended cross-correlation analysis (DCCA) \cite{Podobnik-Stanley-2008-PRL}. Jiang and Zhou extended the multifractal detrending moving-average analysis (MF-DMA) \cite{Gu-Zhou-2010-PRE} and detrending moving-average analysis (DMA)
\cite{Alessio-Carbone-Castelli-Frappietro-2002-EPJB,Carbone-Castelli-2003-SPIE,Carbone-Castelli-Stanley-2004-PRE,Arianos-Carbone-2007-PA,Carbone-2007-PRE,Carbone-Kiyono-2016-PRE,Tsujimoto-Miki-Shimatani-Kiyono-2016-PRE,Kiyono-Tsujimoto-2016-PRE} into the cross multifractal formulism, namely MF-X-DMA \cite{Jiang-Zhou-2011-PRE}. Other multifractal cross-correlation analysis methods include multifractal cross-correlation analysis (MFCCA) \cite{Oswiecimka-Drozdz-Forczek-Jadach-Kwapien-2014-PRE,Kwapien-Oswiecimka-Drozdz-2015-PRE}, and multifractal detrended partial correlation analysis (MFDPXA, including MF-PX-DFA, MF-PX-DMA, and so on) \cite{Qian-Liu-Jiang-Podobnik-Zhou-Stanley-2015-PRE}.

Using these approaches, the long-range cross correlations have been empirically uncovered in pairs of series from different financial markets. Wang {\emph{et al.}} found significant cross correlations between return series of Chinese A-share and B-share markets \cite{Wang-Wei-Wu-2010-PA}. The spot and future markets, like crude oil and CSI 300 index, were reported to exhibit cross multifractal features \cite{Wang-Wei-Wu-2011-PA, Wang-Xie-2013-ND}. Wang and Xie found that the Chinese Currency and four major currencies (USD, EUR, JPY, and KRW)  are significantly cross correlated \cite{Wang-Xie-2013-PA}. Ma {\emph{et al.}} also confirmed the cross correlations between the Chinese stock markets and surrounding stock markets in Japan, South Korea, and Hong Kong \cite{Ma-Wei-Huang-2013-PA}. Wang {\emph{et al.}} report that the returns and trading volumes of CSI 300 index exhibit a long range cross correlated behavior \cite{Wang-Suo-Yu-Lei-2013-PA}. Wang {\emph{et al.}} developed an improved method of minimum-variance hedge ratio, namely the detrended minimum-variance hedge ratio, to capture the hedge ratio at different time scales \cite{Wang-Xie-He-Chen-2014-PA}. Zhou and Chen proposed an arbitrage trading strategy based on the DCCA coefficients and found that this strategy could offer a positive and time-stable return \cite{Zhou-Chen-2016-PA}.

Wavelet transform has long been applied to the study of fractals and multifractals \cite{Holschneider-1988-JSP,Arneodo-Grasseau-Holschneider-1988-PRL} and
a partition function approach based on wavelet transform has been proposed \cite{Muzy-Bacry-Arneodo-1991-PRL}.  Jiang  {\emph{et al.}} generalized the multifractal wavelet analysis to the bivariate case, namely MF-X-WT \cite{Jiang-Zhou-Stanley-2016-PRE}, which is a multifractal generalization of the cross wavelet transform \cite{Hudgins-Friehe-Mayer-1993-PRL,Maraun-Kurths-2004-NPG,AguiarConraria-Soares-2014-JES}. Recently, a new method of wavelet leaders has been proposed to characterize the multifractality \cite{Lashermes-Roux-Abry-Jaffard-2008-EPJB, Serrano-Figliola-2009-PA, Wendt-Roux-Jaffard-Abry-2009-SP}. In this paper, we propose a new joint multifractal analysis based on wavelet leaders, called joint multifractal analysis based on wavelet leaders (MF-X-WL). Similar to the MF-X-PF and MF-X-WT methods, we introduce two orders in the MF-X-WL method. We check the performance of this method by carrying out extensive numerical experiments with two mathematical models and also apply this method to detect the cross multifractality in the pairs of series from financial markets and online worlds.

\section{Methods}
\label{S1:Algo:MF-X-WT}
\subsection{Definition of wavelet leader}

For completeness, we review briefly the definition of wavelet leaders \cite{Lashermes-Roux-Abry-Jaffard-2008-EPJB, Serrano-Figliola-2009-PA, Wendt-Roux-Jaffard-Abry-2009-SP}. Wavelet leaders are defined from the discrete wavelet coefficients, which decompose the signals $x(t)$ on the orthogonal bases  $\{ \psi_{j,k}\}_{j \in Z, k \in Z}$ composed of discrete wavelets $\psi_{j, k}$. Integers $j \in Z$ and $k \in Z$ represent the scale $a = 2^j$ and location $b = k2^j$. Wavelets $\{ \psi_{j,k}\}_{j \in Z, k \in Z}$ are space-shifted and scale-dilated templates of a mother wavelet $\psi_0(t)$, such that,
\begin{equation}
\psi_{j,k}(t) = \frac{1}{2^j} \psi_0 \left(\frac{t-k2^j}{2^j} \right)\label{Eq:MF-X-WL:Basis}
\end{equation}
The mother wavelet $\psi_0(t)$ should have a compact time support and the quadrature mirror filters also have finite impulse responses. In practice, the Daubechies bases are found to satisfy such conditions. In this paper, the Daubechies wavelet with order 1 is used. The discrete wavelet coefficients are defined as follows,
\begin{equation}
d_x(j, k) = \int_t x(t) 2^{-j} \psi_0 (2^{-j} t - k) {\rm{d}}t.  \label{Eq:MF-X-WL:DCT}
\end{equation}

One defines a dyadic interval $\lambda(j, k)$ as
\begin{equation}
\lambda(j,k) = [k2^j, (k+1)2^j),  \label{Eq:MF-X-WL:Lambda:jk}
\end{equation}
and denote the union of the interval $\lambda$ and its 2 adjacent neighbors as $3\lambda$,
\begin{equation}
3\lambda(j,k) = \lambda(j, k-1) \cup \lambda(j, k) \cup \lambda(j, k+1) ,  \label{Eq:MF-X-WL:3Lambda:jk}
\end{equation}
Following Ref.~\cite{Lashermes-Roux-Abry-Jaffard-2008-EPJB}, the wavelet leader $L_x(j, k)$ is defined as
\begin{equation}
L_x(j, k) =  \sup_{\lambda' \subset 3\lambda(j, k) } |d_x(\lambda')|,  \label{Eq:MF-X-WL:WL}
\end{equation}
The physical meaning of Eq.~(\ref{Eq:MF-X-WL:WL}) is that the wavelet leader $L_x(j, k)$ corresponds to the largest value of the absolute wavelet coefficients $|d_{x}(j',k')|$ calculated on intervals, $(k-1)2^j \le 2^{j'} k' < (k+2)2^j$, with $0 < j' \le j$. Note that all the fine scales $2^{j'} \le 2^j$ must be considered to compute the wavelet leaders.

\subsection{Cross multifractal formalism based on wavelet leaders}

Motivated from the multifractal formalism of wavelet leaders \cite{Lashermes-Roux-Abry-Jaffard-2008-EPJB, Serrano-Figliola-2009-PA, Wendt-Roux-Jaffard-Abry-2009-SP} and the multifractal cross correlation analysis \cite{Podobnik-Stanley-2008-PRL, Zhou-2008-PRE, Jiang-Zhou-2011-PRE, Kristoufek-2011-EPL, Wang-Shang-Ge-2012-Fractals, Xie-Jiang-Gu-Xiong-Zhou-2015-NJP}, we propose an algorithm to detect the cross multifractality in a pair of series, $x(t)$ and $y(t)$, based on wavelet leaders, namely joint multifractal analysis based on wavelet leaders with two moment orders $p$ and $q$ (MF-X-WL $(p, q)$). Firstly, the wavelet leaders of both series are estimated at different scales $a = 2^j$, giving $L_x(j, k)$ and $L_y(j, k)$. For a given scale $2^j$, we can define the joint partition function with moment orders $p$ and $q$ based on wavelet leaders,
\begin{equation}
S^L_{xy}(p, q, j) = \frac{1}{n_j}\sum_{k=1}^{n_j} L_x(j,k)^{p/2}L_y(j,k)^{q/2}, \label{Eq:MF-X-WL:SF}
\end{equation}
where $n_j$ is the number of wavelet leaders at scale $2^j$. When $x = y$ and $p=q$, we recover the traditional multifractal formalism based on wavelet leaders. One can also expect the following scaling behavior if the underlying processes are cross multifractal,
\begin{equation}
S^L_{xy}(p, q, j) \sim 2^{j \zeta^L_{xy}(p, q)}.\label{Eq:MF-X-WL:SF:Scaling}
\end{equation}
where $\zeta^L_{xy}(p, q)$ is the joint scaling exponents. Obviously, we can estimate  $\zeta^L_{xy}(p, q)$ through regressing $\ln \chi_{xy} (p, q, j)$ against $j \ln 2$ in the scaling range for a given pair $(p,q)$.

In analogy with the double Legendre transforms in MF-X-PF $(p,q)$ \cite{Xie-Jiang-Gu-Xiong-Zhou-2015-NJP} and the multifractal formalism of wavelet leaders \cite{Lashermes-Roux-Abry-Jaffard-2008-EPJB, Serrano-Figliola-2009-PA, Wendt-Roux-Jaffard-Abry-2009-SP}, we can obtain singularity strengths $h_x$ and $h_y$
\begin{eqnarray}
 h_x(p,q) & = & 2\partial \zeta^L_{xy}(p, q) / \partial p, \label{Eq:MF-X-WL:hx}\\
 h_y(p,q) & = & 2\partial \zeta^L_{xy}(p, q) / \partial q, \label{Eq:MF-X-WL:hy}
\end{eqnarray}
and the multifractal spectrum $D_{xy}(h_x, h_y)$ for MF-X-WL $(p,q)$
\begin{equation}
 D_{xy}(h_x, h_y) = 1 + p h_x /2 + q h_y /2  - \zeta^L_{xy}. \label{Eq:MF-X-WL:Dh}
\end{equation}

As pointed out by Muzy {\emph{et al}}. \cite{Muzy-Bacry-Arneodo-1991-PRL}, the estimation of singularity strength and multifractal spectrum based on the Legendre transform may have various errors because of its innate disadvantages. They also proposed an alternative method to compute $h$ and $D(h)$ from the perspective of canonical method, which is known as a direct estimation method \cite{Chhabra-Jensen-1989-PRL}. This inspires us to directly estimate the singularity strength $h_x$ and $h_y$ and the multifractal spectrum $D_{xy}(p, q)$ through the following equations,
\begin{eqnarray}
 h_x(p,q) & = & \lim_{s \rightarrow 0} \frac{1}{ \ln 2^j } \sum_k \mu_{xy}(p,q,j,k) \ln L_x(j, k), \label{Eq:MF-X-WL:DE:hx}\\
 h_y(p,q) & = & \lim_{s \rightarrow 0} \frac{1}{ \ln 2^j } \sum_k \mu_{xy}(p,q,j,k) \ln L_y(j, k), \label{Eq:MF-X-WL:DE:hy} \\
 D_{xy}(p,q) &=& \lim_{s \rightarrow 0} \frac{1}{\ln 2^j} \sum_k \mu_{xy}(p,q,j,k) \ln \mu_{xy}(p,q,j,k). \label{Eq:MF-X-WL:DE:Dxy}
\end{eqnarray}
where $\mu_{xy} (p,q,j,k) = \frac{L_x(j,k)^{p/2} L_y(j,k)^{q/2}}{\sum_k L_x(j,k)^{p/2} L_y(j,k)^{q/2}}$. Thus we can directly determine the singularity strength, $h_x(p,q)$ and $h_y(p,q)$, and multifractal function, $D_{xy}(p,q)$, from log-log plots of the quantities in Eqs.~(\ref{Eq:MF-X-WL:DE:hx})--(\ref{Eq:MF-X-WL:DE:Dxy}).

\section{Numerical experiments}
\label{S1:NumSim}

Here, we first conduct two numerical experiments, including binomial measures generated from the multiplicative $p$-model \cite{Meneveau-Sreenivasan-1987-PRL} and bivariate fractional Brownian motions (bFBMs) \cite{Lavancier-Philippe-Surgailis-2009-SPL,Coeurjolly-Amblard-Achard-2010-EUSIPCO,Amblard-Coeurjolly-Lavancier-Philippe-2013-BSMF}, to test the validity and performance of the proposed MF-X-WL $(p,q)$ approach.

\subsection{Joint multifractal analysis of binomial measures}

\begin{figure*}[htb]
\centering
\includegraphics[width=15cm]{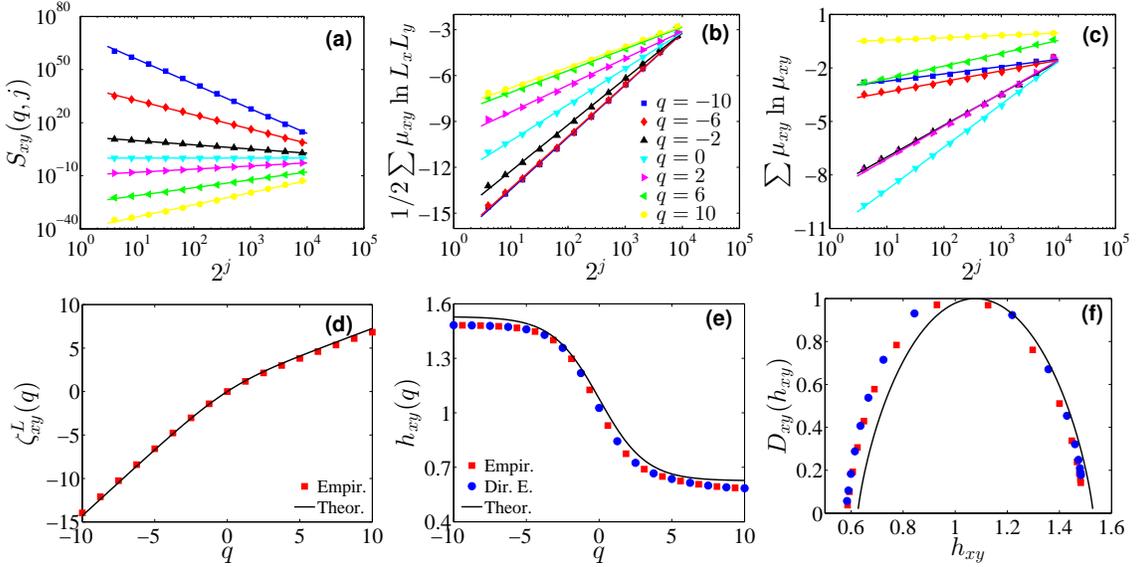}
\caption{\label{Fig:MF-X-WL:pModel:XY:Q} (Color online) Joint multifractal analysis of two binomial measures with $p_x=0.3$ and $p_y = 0.4$ based on the MF-X-WL($q$) method.  (a) Power-law behaviors between $S_{xy}(q, j)$ and the scale $2^j$ for different $q$ values. (b) Plots of $\sum_k \mu_{xy}(q,j,k) \ln L_x(j,k) L_y(j,k)^{1/2}$ against $2^j$. (c) Plots of $\sum_k \mu_{xy} (q,j,k) \ln \mu_{xy} (q,j,k)$ with respect to $2^j$. (d) Scaling exponent function $\zeta^L_{xy}(q)$. (e) Singularity strength function $h_{xy}(q)$. (f) Multifractal singularity spectrum $D_{xy}(h_{xy})$.}
\end{figure*}

As a first example, we conduct a numerical experiment of testing the validity of our algorithm on two binomial measures $\{x(t): t = 1, 2, \cdots, 2^j\}$ and $\{y(t): t = 1, 2, \cdots, 2^j\}$ from the $p$-model with known analytic multifractal properties \cite{Meneveau-Sreenivasan-1987-PRL}. Each binomial measure is generated in an iterative manner. We start with the zeroth iteration $i = 0$, where the data set $z(i)$ consists of one value, $z^{(0)}(1)= 1 $. In the $i$-th iteration, the data set $\{z^{(i)}(t): t = 1, 2, \cdots, 2^j\}$ is obtained from
\begin{equation}
  \begin{array}{l}
    z^{(i)}(2j-1)= p_z z^{(i-1)}(j),\\
    z^{(i)}(2j)  = (1-p_z)z^{(i-1)}(j),
\end{array}
  \label{Eq:pModel}
\end{equation}
for $i = 1, 2, \cdots, 2^{j-1}$. When $i\to\infty$, $z^{(i)}(t)$ approaches to a binomial measure, whose partition function $S_{zz}(q, j)$ and scaling exponent function $ \zeta_{zz}(q) $ have an analytic form \cite{Serrano-Figliola-2009-PA},
\begin{eqnarray}
 S_{zz}(q, j) & = & 2^{-j} (p_z^q + (1-p_z)^q)^j, \label{Eq:pModel:Szz}\\
\zeta_{zz}(q) & = & 1 -\log_2[p_z^q+(1-p_z)^q]. \label{Eq:pModel:Zetazz}
\end{eqnarray}
In our numerical experiments, the $p$-model parameters of two binomial measures are set as $p_z=p_x =0.3$ for $x(i)$ and $p_z=p_y = 0.4$ for $y(i)$ with an iterative number $i=16$. The analytic scaling exponent functions $\zeta_{xx}(q)$ and $\zeta_{yy}(q)$ of $x$ and $y$ are expressed in Eq.~(\ref{Eq:pModel:Szz}), when we replace $z$ with $x$ and $y$. Because both series are generated in terms of the same rule, the two series $x$ and $y$ are strongly correlated with a coefficient of 0.82. In the case of cross multifractal analysis, we have the theoretical expressions for the partition function and scaling exponent function,
\begin{eqnarray}
 S_{xy}(p, q, j) & = & 2^{-j} [p_x^{p/2}p_y^{q/2} + (1-p_x)^{p/2}(1-p_y)^{q/2}]^j, \label{Eq:pModel:Sxy}\\
\zeta_{xy}(p, q) & = & 1 -\log_2[p_x^{p/2}p_y^{q/2} + (1-p_x)^{p/2}(1-p_y)^{q/2}]. \label{Eq:pModel:Zetaxy}
\end{eqnarray}
We first consider the scenario of $p=q$. The results are shown in Fig.~\ref{Fig:MF-X-WL:pModel:XY:Q}. 

Fig.~\ref{Fig:MF-X-WL:pModel:XY:Q} (a) illustrates the power-law behaviors, spanning more than three orders of magnitude, between the partition functions $S_{xy}(q,j)$ and the scale $2^j$. Fig.~\ref{Fig:MF-X-WL:pModel:XY:Q} (b) and (c) present the linear behaviors of the two quantities $\sum \mu_{xy} \ln (L_xL_y)^{1/2}$ and $\sum \mu_{xy} \ln \mu_{xy}$ against $\ln (2^j)$. By linearly regressing the data in plots (a) - (c), we can obtain the scaling exponent function $\zeta^L(q)$, the singularity strength $h_{xy}(q)$, and the multifractal function $D_{xy}(q)$. In Fig.~\ref{Fig:MF-X-WL:pModel:XY:Q} (d), the estimated scaling exponents $\zeta^L(q)$ and theoretical function $\zeta(q)$ in Eq.~(\ref{Eq:pModel:Zetaxy}) are plotted with respect to $q$ for comparison. One can see that the empirical and theoretical values agree with each other nicely when $q \le 5$ and the estimation errors are also acceptable when $q>5$ , suggesting that MF-X-WL$(q)$ has a good performance in detecting the cross multifractal nature in two binomial measures. Furthermore, the nonlinear behavior between $\zeta(q)$ and $q$ is a hallmark of multifractality, agreeing with our expectation. Fig.~\ref{Fig:MF-X-WL:pModel:XY:Q} (e) presents a comparison of the singularity strength $h_{xy}(q)$ obtained from different methods. The solid line corresponds to the theoretical values. The squares and circles are obtained from the first derivation of the scaling exponents $\zeta^L_{xy}(q)$ and the estimation of the slopes in Fig.~\ref{Fig:MF-X-WL:pModel:XY:Q} (b). One can find that the square and circle curve perfectly coincide with each other. However, both curves exhibit a downward shift from the theoretical line. Fig.~\ref{Fig:MF-X-WL:pModel:XY:Q} (f) illustrates the multifractal spectra of two binomial measures, in which a theoretical line and two estimated curves are plotted. Both empirical curves $D_{xy}(h_xy)$, obtained from Eq.~(\ref{Eq:MF-X-WL:Dh}) (squares) and Eq.~(\ref{Eq:MF-X-WL:DE:Dxy}) (circles), are again in good agreement with each other and exhibit deviations from the theoretical line on the left side, resulting from the estimation errors in $\zeta_{xy}(q)$ when $q > 5$. Our results suggest that the MF-X-WL $(q)$ is able to provide acceptable results in the analysis of cross mutifractality in two binomial measures.

\begin{figure*}[htb]
\centering
\includegraphics[width=18cm]{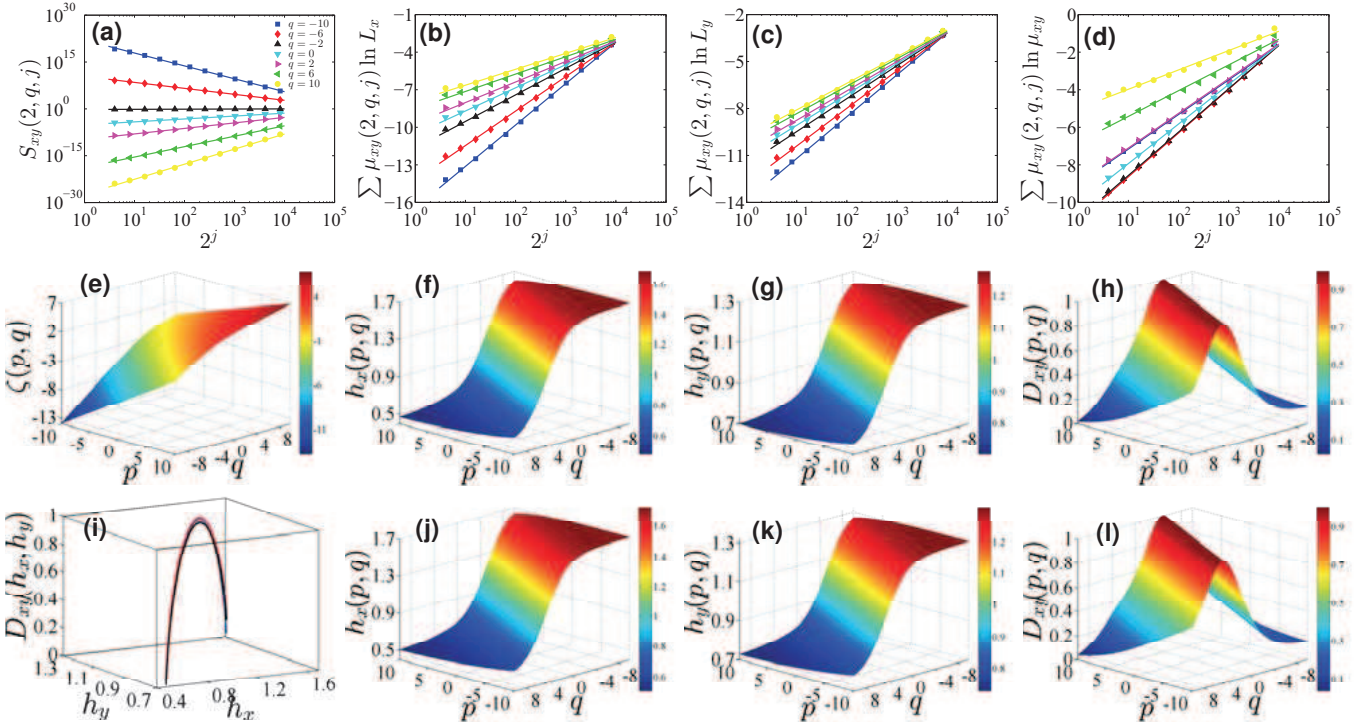}
\caption{\label{Fig:MF-X-WL:pmodel:XY:PQ} (Color online) Joint multifractal analysis of two binomial measures with $p_x = 0.3$ and $p_y = 0.4$ based on the MF-X-WL$(p,q)$ method. (a) Power-law plots of $S_{xy}(p,q,j)$ with respect to the scale $2^j$ for different $q$ with fixed $p=2$. (b) Plots of $\sum_t \mu_{xy} (2,q,j,k) \ln L_x(j,k)$ against $2^j$ for different $q$ with fixed $q=2$. (c) Plots of $\sum_t \mu_{xy} (2,q,j,k) \ln L_y(j,k)$ against $2^j$ for different $q$ with fixed $q=2$. (d) Plots of $\sum_t \mu_{xy} (2,q,j,k) \ln \mu_{xy} (2,q,j,k)$ against $2^j$ for different $q$ with fixed $q=2$. (e)--(h) Mass exponent function $\zeta_{xy} (p,q)$, singularity functions $h_x(p,q)$ and $h_y(p,q)$, and multifractal function $D_{xy}(p,q)$ from Eqs.~(\ref{Eq:MF-X-WL:SF:Scaling})--(\ref{Eq:MF-X-WL:Dh}). (i) Mulfractal spectra $D_{xy}(h_x, h_y)$. (j)--(l) Singularity functions $h_x(p,q)$ and $h_y(p,q)$ and multifractal function $D_{xy}(p,q)$ from Eqs.~(\ref{Eq:MF-X-WL:DE:hx})--(\ref{Eq:MF-X-WL:DE:Dxy}). }
\end{figure*}

We then release the restriction of $p=q$. Fig.~\ref{Fig:MF-X-WL:pmodel:XY:PQ} illustrate the corresponding results. In Fig.~\ref{Fig:MF-X-WL:pmodel:XY:PQ} (a), we present the power-law dependence between the partition function $S_{xy} (2,q,j)$ and the scale $2^j$ for different $q$ with fixed $p=2$. The power-law behavior spans more than three orders of magnitudes. Fig.~\ref{Fig:MF-X-WL:pmodel:XY:PQ} (b) -- (d) illustrates the plots of the three quantities $\sum \mu_{xy} \ln L_x$, $\sum \mu_{xy} \ln L_y$, and $\sum \mu_{xy} \ln \mu_{xy}$ with respect to the scale $2^j$. Again, very nice linear behaviors are observed between the three quantities and the logarithmic scale. The power-law exponents in panel (a) correspond to the scaling exponents $\zeta_{xy}(p,q)$. In Fig.~\ref{Fig:MF-X-WL:pmodel:XY:PQ} (e), we plot the scaling exponents $\zeta_{xy}(p,q)$ as a function of $p$ and $q$. Obviously, we can see that $\zeta_{xy}(p,q)$ is a nonlinear function of $p$ and $q$, verifying the cross multifractility in the two binomial measures.

Through the double Legendre transform presented in Eqs.~(\ref{Eq:MF-X-WL:hx})-(\ref{Eq:MF-X-WL:Dh}), we can determine numerically the two singularity strength functions $h_x(p,q)$ and $h_y(p,q)$ and the multifractal spectrum $D_{xy}(h_x, h_y)$ from $\zeta_{xy}(p, q)$. The corresponding $h_x(p,q)$, $h_y(p,q)$ and $D_{xy}(p,q)$ are shown in Fig.~\ref{Fig:MF-X-WL:pmodel:XY:PQ} (f), (g), and (h), respectively. Alternatively, Eqs.~(\ref{Eq:MF-X-WL:DE:hx})-(\ref{Eq:MF-X-WL:DE:Dxy}) provide another way to directly estimate the joint singularity strengthes $h_x$ and $h_y$ and the multifractal function $D_{xy}(p, q)$, which are the slopes in panels (b), (c), and (d). And the results from the direct methods are illustrated in Fig.~\ref{Fig:MF-X-WL:pmodel:XY:PQ} (j), (k), and (l). In Fig.~\ref{Fig:MF-X-WL:pmodel:XY:PQ} (i), we plot the theoretical multifracal spectrum (blue dots) and two empirical multifractal spectra. One of the empirical spectra is obtained from the Legendre transform of $\zeta_{xy}$ (black dots) and the other empirical spectrum is given by the direct determination approach (red dots). We can find that the three multifractal spectra $D_{xy} (h_x, h_y)$ are not a planar surface, but a surface with curvatures, suggesting the univariate function relationship between $D_{xy}$ and $h_x$ and/or $h_y$. Such univariate function behavior of multifractal spectra is also uncovered by the MF-X-PF$(p, q)$ method \cite{Xie-Jiang-Gu-Xiong-Zhou-2015-NJP}. The two empirical multifractal spectra do not overlap exactly with the theoretical spectrum, suggesting the existence of estimation errors when applying MF-X-WL $(p,q)$ to test the joint multifractal nature of two binomial measures.

\subsection{Joint multifractal analysis of bivariate fractional Brownian motions}

A bivariate fractional Brownian motion (bFBM) $[x(t),y(t)]$ with parameters $\{H_{xx},H_{yy}\}\in(0,1)^2$ is a self-similar Gaussian process with stationary increments, where $x(t)$ and $y(t)$ are two univariate fractional Brownian motions with Hurst indices $H_{xx}$ and $H_{yy}$ and are the two components of the bFBM and the basic properties of multivariate fractional Brownian motions have been extensively studied \cite{Lavancier-Philippe-Surgailis-2009-SPL,Coeurjolly-Amblard-Achard-2010-EUSIPCO,Amblard-Coeurjolly-Lavancier-Philippe-2013-BSMF}. Extensive numerical experiments of multifractal cross-correlation analysis algorithms have been performed on bFBMs \cite{Jiang-Zhou-2011-PRE, Qian-Liu-Jiang-Podobnik-Zhou-Stanley-2015-PRE, Xie-Jiang-Gu-Xiong-Zhou-2015-NJP}. The two Hurst indexes $H_{xx}$ and $H_{yy}$ of the two univariate FBMs and their cross-correlation coefficient $\rho$ are input arguments of the bFBM synthetic algorithm. By using the simulation procedure describe in Ref.~\cite{Coeurjolly-Amblard-Achard-2010-EUSIPCO, Amblard-Coeurjolly-Lavancier-Philippe-2013-BSMF}, we have generated a realization of bFBM with $H_{xx} = 0.5$, $H_{yy} = 0.8$, and $\rho=0.3$. The length of the bFBM is $2^{16}$.

\begin{figure*}[htb]
\centering
\includegraphics[width=15cm]{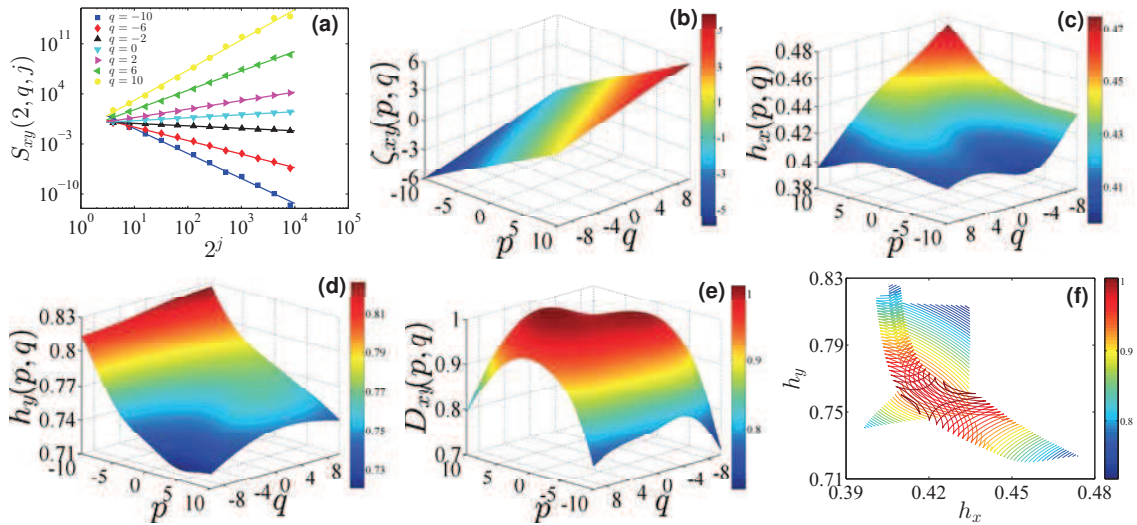}
\caption{\label{Fig:MF-X-WL:BFBM} (Color online) Joint multifractal analysis of bivariate fractional Brownian motion with $H_{xx} = 0.5$, $H_{yy} = 0.8$, and $\rho = 0.3$. (a) Power-law relationship between $S_{xy}(p, q, j)$ and scale $2^j$ for different $q$ with fixed $p=2$. (b) Scaling exponent function $\zeta_{xy}(p,q)$ obtained from Eq.~(\ref{Eq:MF-X-WL:SF:Scaling}). The linear least-squares regression shows that $\zeta_{xy} = 0.2112p+0.3809q-0.0168$. (c) Singularity function $h_x(p, q)$. (d) Singularity function $h_y(p, q)$. (e) Multifractal spectrum $D_{xy}(p,q)$. (f) Contour plots of multifractal spectrum $D_{xy}(h_x, h_y)$.}
\end{figure*}

Following Ref.~\cite{Xie-Jiang-Gu-Xiong-Zhou-2015-NJP}, the cross singularity strength $h_{xy}$ is a constant if each component is monofractal, which also leads to the constant cross multifractal spectrum $D_{xy}(h_x, h_y)$. In Fig.~\ref{Fig:MF-X-WL:BFBM}, we present the results of the joint multifractal analysis on bFBMs using the MF-X-WL$(p,q)$ algorithm. Fig.~\ref{Fig:MF-X-WL:BFBM} (a) illustrates the joint partition function $S_{xy}(2,q,s)$ against the scale $2^j$ for different $q$ with fixed $p=2$. Intriguing power-law behaviors are observed between the partition functions and the scales. The least-squares estimation gives the scaling exponents $\zeta_{xy}$. In Fig.~\ref{Fig:MF-X-WL:BFBM} (b), the scaling exponents $\zeta_{xy}$ are plotted with respect to the moment orders $p$ and $q$. A plane is observed, indicating that there is a linear relationship between $\zeta_{xy}$ and the moment orders $p$ and $q$. Such result agrees well with the theoretical expectation. The bivariate regression gives that
\begin{equation}
  \zeta_{xy}(p, q) = 0.2112 p + 0.3809q -0.0168.
  \label{Eq:MFXWL:bFBM:Zeta:p:q} 
\end{equation}
Comparing with Eq.~(\ref{Eq:MF-X-WL:Dh}) we have ${\overline{h}_x} = 0.4224$, ${\overline{h}}_y = 0.7618$, and ${\overline{D}}_{xy} = 1.0168$. 

Eqs.~(\ref{Eq:MF-X-WL:hx}) and (\ref{Eq:MF-X-WL:hy}) also provide an approximate way to estimate $h_x$ and $h_y$. In Fig.~\ref{Fig:MF-X-WL:BFBM} (c) and (d), we plot the joint singularity strength $h_x$ and $h_y$, numerically estimated from taking the forward difference of ${\zeta}_{xy}$, with respect to the moment orders $p$ and $q$. The estimated joint singularity strengths $h_x$ and $h_y$ vary in a very narrow range, verifying the monofractal features. The average values of $h_x$ and $h_y$ are $0.4229$ and $0.7652$, which nicely agree with  ${\overline{h}}_x$ and ${\overline{h}}_y$. Based on $h_x$ and $h_y$, we can further give the multifractal function following Eq.~(\ref{Eq:MF-X-WL:Dh}), which is plotted with respect to $p$ and $q$ in Fig.~\ref{Fig:MF-X-WL:BFBM} (e), and against $h_x$ and $h_y$ in Fig.~\ref{Fig:MF-X-WL:BFBM} (f). The singularity function $D_{xy}$ varies in a range from 0.7 to 1 with a mean value of 0.9457, smaller than ${\overline{D}}_{xy} =1.0168$. The estimated $h_x$, $h_y$, and $D_{xy}$ from both approaches differ to certain degree from the corresponding theoretical values of $h_{xx} = 0.5$, $h_{yy} = 0.8$, and $D_{xy} = 1$. The wide spanning range of $D_{xy}$ also indicates that MF-X-WL may give spurious multifractality for bFBMs. The spurious results also strength the necessity of performing statistical tests on checking multifractality based on bootstrapping \cite{Jiang-Zhou-2007-PA, Jiang-Xie-Zhou-2014-PA}.

\section{Applications}
\label{S1:Application}

\subsection{Financial markets}

In this section, we first apply the MF-X-WL $(p,q)$ algorithm to uncover the cross multifractality in daily returns and volatilities of the Dow Jones Industrial Average (DJIA) index and the National Association of Securities Dealers Automated Quotations (NASDAQ) index. The daily return is defined as the logarithmic difference of daily closing price:
\begin{equation}
  R(t) = \ln I(t) - \ln I(t-1), 
  \label{Eq:MF-X-WT:FinIndex:Return}
\end{equation}
where $I(t)$ is the closing price of the DJIA or the NASDAQ on day $t$. Both indexes are retrieved from $!$Yahoo finance. The spanning period of both indexes is from 5 December 1983 to 17 June 2016, containing 8192 data points in total. The volatilities are defined as the absolute values of the daily returns.

\begin{figure*}[htb]
  \centering
  \includegraphics[width=18cm]{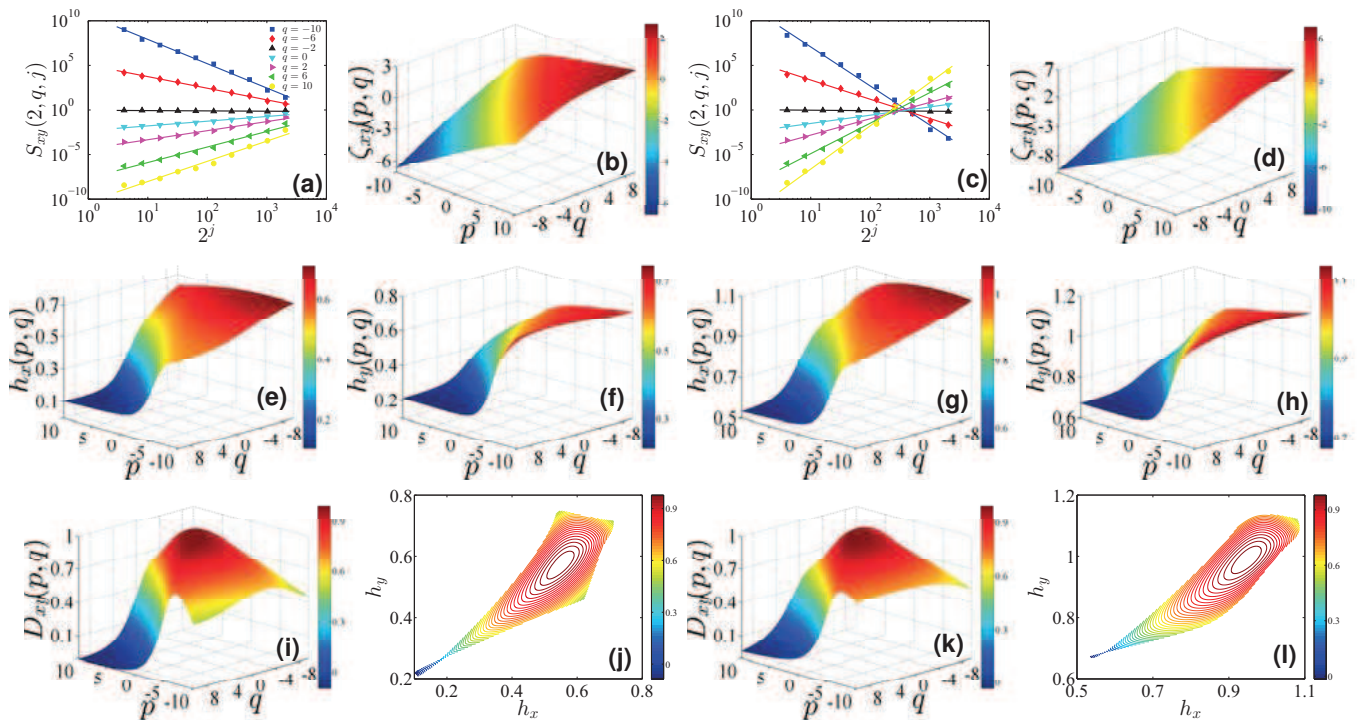}
  \caption{\label{Fig:MF-X-WL:FM} (Color online) Empirical cross multifractal analysis in financial markets using the MF-X-WL$(p,q)$ algorithm. (Left two columns) Cross multifractality between the daily returns of the DJIA index and the NASDAQ index. (Right two columns)  Cross multifractality between the daily volatilities of the DJIA index and the NASDAQ index. (a, c) Power-law plots of $S_{xy}(p, q, j)$ with respect to the scale $2^j$ for different $q$ with fixed $p=2$.  (b, d) Mass exponent function $\zeta_{xy}(p,q)$. (e, g) Singularity strength function $h_x(p,q)$. (f, h) Singularity strength function $h_y(p,q)$. (i, k) Multifractal function $D_{xy} (p,q)$. (j, l) Multifractal singularity spectrum $D_{xy}(h_x, h_y)$. }
\end{figure*}

We present the empirical results when applying MF-X-WL$(p, q)$ on the returns and volatilities of the DJIA index and the NASDAQ index in Fig~\ref{Fig:MF-X-WL:FM}. The left two columns and right two columns of Fig.~\ref{Fig:MF-X-WL:FM} correspond to the cross multifractal analysis of the two return series and the two volatility series, respectively. 

In panels (a) and (c), the partition function $S_{xy}(2,q,j)$ is plotted with respect to the scale $2^j$ for different $q$ with fixed $p=2$. One can see that in both panels the power-law scaling is extremely good, spanning over two orders of magnitude. The linear regression of $\ln S(p,q,j)$ with respect to $\ln 2^j$ for a given pair of $p$ and $q$, gives the mass exponents $\zeta_{xy}(p,q)$, which are shown in panels (b) and (d). Intriguingly, the mass exponents are a nonlinear function of $p$ and $q$ in the two panels, indicating the presence of cross multifractal features in the return series and the volatility series. 

The joint singularity strength functions $h_x$ and $h_y$ are plotted in panels (e, g) and panels (f, h), respectively. The singularity functions are numerically estimated from $\zeta_{xy}(p,q)$. We note that the singularity functions $h_x$ and $h_y$ of returns and volatilities are well dispersed with the spanning ranges greater than 0.5 and are a monotonic function of $p$ or $q$. Such a width of both singularity strength functions confirms the existence of joint multifractality in the two return series and volatility series. We also find that the joint singularities of volatilities are greater than those of returns on average, indicating stronger memory behaviors in the volatility pairs.

Fig.~\ref{Fig:MF-X-WL:FM} (i) and (k) illustrate the multifractal function $D_{xy}(p,q)$ obtained from the double Legendre transform. It is observed that the multifractal function locates in the range of $(0, 1)$ with the maximum point at the point of $(0,0)$ in both panels. In Fig.~\ref{Fig:MF-X-WL:FM} (j) and (l), we sketch the multifractal spectrum $D_{xy}(h_x, h_y)$ for returns and volatilities. Our empirical findings favor the existence of joint multifractality in return and volatility pairs of DJIA and NASDAQ.

\subsection{Online world}

Following Ref.~\cite{Jiang-Ren-Gu-Tan-Zhou-2010-PA}, we next investigate the cross multifractal feature in the online number of avatars in a massive multiplayer online role-playing game (MMORPG). Our online number records how many avatars are online simultaneously in each minute. We extract two pairs of online number of avatars from a popular MMORPG in China. In this virtual world, players need to select gender when they create their avatars. And when avatars achieve the level of 16, their operators need to assign the avatars to be a member in one of two opposed societies, namely, Xian and Mo societies. The first pair corresponds to the online number of male $n_m$ and female $n_f$ avatars. The second pair is the online number of avatars in Xian $n_X$ and Mo $n_M$ societies. Each series covers a period from June 2, 2011 to September 3, 2011 with a total number of 131072 points. We then apply the MF-X-WL$(p,q)$ to analyze the cross multifractal nature in each pair of the online numbers.

\begin{figure*}[htb]
\centering
\includegraphics[width=18cm]{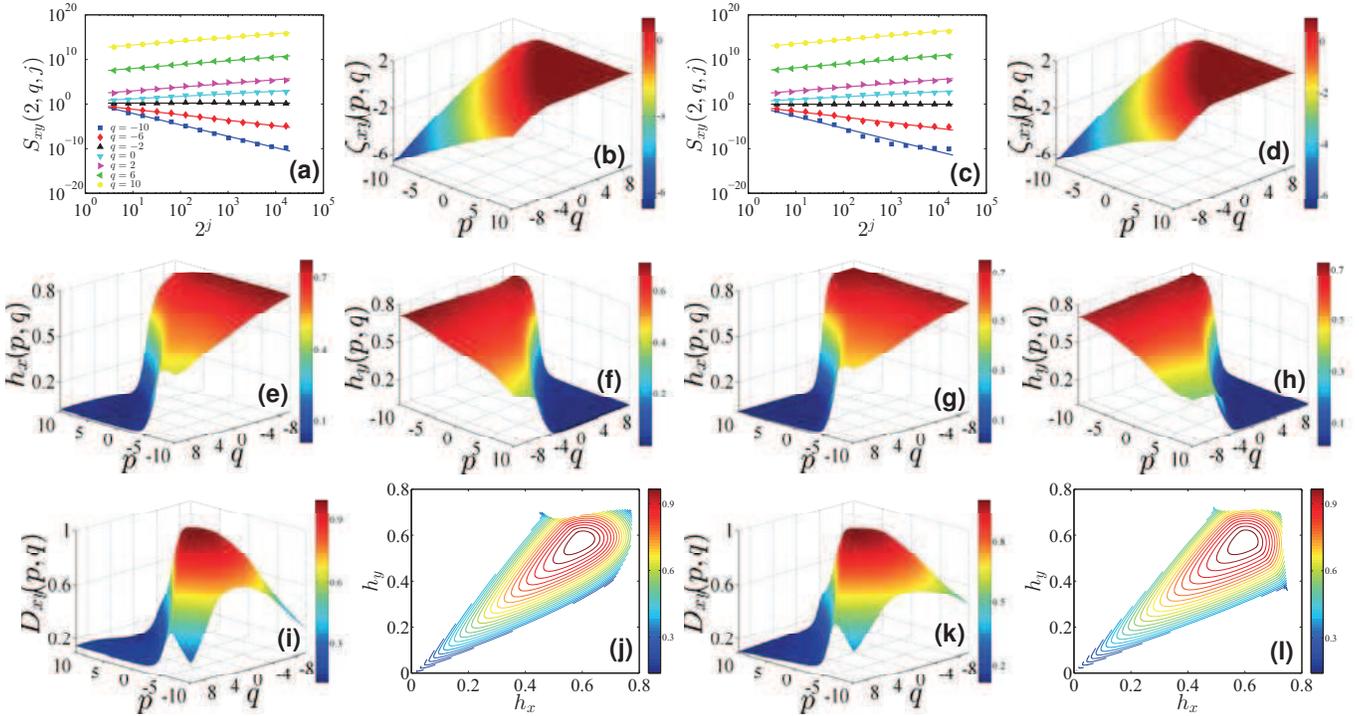}
\caption{\label{Fig:MF-X-WL:OW} (Color online) Empirical cross multifractal analysis in an online world using the MF-X-WL$(p,q)$ algorithm. (Left two columns) Cross multifractality between between the online number of male avatars $n_m$ and the number of female avatars $n_f$. (Right two columns)  Cross multifractality between  the online number of avatars in Xian society $n_X$ and the number of avatars in Mo society $n_M$.  (a, c) Power-law plots of $S_{xy}(p, q, j)$ with respect to the scale $2^j$ for different $q$ with fixed $p=2$.  (b, d) Mass exponent function $\zeta_{xy}(p,q)$. (e, g) Singularity strength function $h_x(p,q)$. (f, h) Singularity strength function $h_y(p,q)$. (i. k) Multifractal function $D_{xy} (p,q)$. (j, l) Multifractal singularity spectrum $D_{xy}(h_x, h_y)$. }
\end{figure*}

The empirical results of cross multifractal analysis are plotted in Fig.~\ref{Fig:MF-X-WL:OW} for two pairs of online numbers. We show the empirical cross multifractality in the pair of $n_m$ and $n_f$ in the left two columns and of $n_X$ and $n_M$ in the right two columns. 

Fig.~\ref{Fig:MF-X-WL:OW} (a) and (c) illustrate the power-law relationship between the partition function $S_{xy}(2,q,j)$ and the scale $2^j$ for different $q$ with fixes $p=2$. We observe that the spanning range of the power-law is more than three orders of magnitude. The mass exponents $\zeta_{xy}(p,q)$, corresponding the power-law exponents of $S_{xy}$ versus $2^j$, are shown in Fig.~\ref{Fig:MF-X-WL:OW} (b) and (d). In both panels, the mass exponents increase nonlinearly with respect to $p$ or $q$, implying the evident cross multifractality in both pairs of the online numbers. 

Following Eqs.~(\ref{Eq:MF-X-WL:hx}-\ref{Eq:MF-X-WL:hy}), we numerically estimate the joint singularity strength  functions $h_x$ and $h_y$ and plot them in Fig.~\ref{Fig:MF-X-WL:OW} (e, g) and (f, h), respectively. One can see that, for both pairs of online numbers, the shapes of $h_x$ are very similar to each other. We also observe that the maximum value, the spanning range, and the minimum value of $h_x$ are all very close in panels (e) and (g). For the joint singularity strength $h_y$, very similar behaviors are observed. The widths of both singularity strength functions are very close to 0.8, further confirming the existence of cross multifractal nature in both pairs of online numbers. In Fig.~\ref{Fig:MF-X-WL:OW} (i, k) and (j, l), we further plot the multifractal function $D_{xy}(p,q)$ and the multifractal spectrum $D_{xy}(h_x, h_y)$ for the pair of $n_m$ and $n_f$ and the pair of $n_X$ and $n_M$, respectively. The wide spanning range of $D_{xy}$ again offer strong evidence in favor of the multifractal characteristics in the cross-correlation in both pairs of online numbers.

\section{Summary and conclusions}
\label{S1:conclusion}

We have developed a new approach for joint multifractal analysis with two moment orders based on wavelet leaders, termed as MF-X-WL$(p,q)$. The MF-X-WL approach overcomes the shortcoming of the multifractal wavelet analysis that the moment order much be positive. Extensive numerical experiments has been carried out to check the performances of the MF-X-WL$(p,q)$ method, in which the testing time series pairs are generated from binomial measures and bivariate fractional Brownian motions. Furthermore, this MF-X-WL$(p,q)$ method has also been applied to the time series pairs from financial markets and online worlds to test its ability to detect any joint multifractalities, in which the testing data include pairs of returns, volatilities, and online avatar numbers.

In the numerical experiments, we found that the MF-X-WL$(p,q)$ method is able to detect respectively the joint multifractality and monofractality in binomial measures and bivariate  fractional Brownian motions, but is not able to give very accurate results. For the synthetic data, we can obtain very nice power-law scaling behaviors between the partition functions and the scale. And we can verify the nonlinear (respectively, linear) feature of the scaling exponent $\zeta_{xy}$ against $p$ or $q$ for binomial measures (bivariate fractional Brownian motions). However, the estimation errors are propagated when we calculate the joint singularity strengths and the multifractal functions (see Fig.~\ref{Fig:MF-X-WL:pModel:XY:Q} (e, f), Fig.~\ref{Fig:MF-X-WL:pmodel:XY:PQ} (i) and Fig.~\ref{Fig:MF-X-WL:BFBM} (c- f)). As we know, comparing with the DFA-based methods, the wavelet based methods are not the best approach for multifractal analysis \cite{Kantelhardt-Zschiegner-KoscielnyBunde-Havlin-Bunde-Stanley-2002-PA, Oswiecimka-Kwapien-Drozdz-2006-PRE}. Possible explanations are as follows: (1) Many other methods, like DFA, are directly applied on the series itself, while the wavelet-based approach is performed on the wavelet coefficients of the analyzed series. The wavelet transform could introduce computational errors, which could be magnified in the followup multifractal analysis. (2) The wavelet leaders are the maximum absolute wavelet coefficients across all scales under consideration. Dropping the non-maximum wavelet coefficients may induce errors in the followup analysis, because these values may have the information of multifractality. However, it is still worth to investigate wavelet-based multifractal methods, since wavelet analysis is wildly accepted in the field of economics and finance, compared to other Econophysics methods.

In the applications, we have used the MF-X-WL method to analyze the joint multifractality in pairs of returns and volatilities in stock markets and pairs of online avatar numbers in an online world. All the empirical results confirm the presence of joint multifractality in pairs of financial series and online number series. Because of lacking theoretical solutions of multifractal formulism for real world series, we cannot offer the impression about the accuracy of the MF-X-WL$(p,q)$. One possible solution is to compare with the results of other methods, like MF-X-DFA and MF-X-PF, which is however for future research.

\begin{acknowledgments}
We acknowledge financial support from the National Natural Science Foundation of China (11375064 and 71131007), the Program for Changjiang Scholars and Innovative Research Team in University (IRT1028), and the Fundamental Research Funds for the Central Universities.
\end{acknowledgments}

\bibliography{E:/Papers/Auxiliary/Bibliography}

\end{document}